\documentclass[aps,pre,preprint,superscriptaddress,showpacs]{revtex4}
\usepackage{amssymb}
\usepackage{amsmath}
\usepackage{amsfonts}
\usepackage{graphicx}

\begin{document}

\title{Global Spatiotemporal Order and Induced Stochastic Resonance
due to a Locally Applied Signal}
\author{A.~Samoletov}
\affiliation{The SIMBIOS Centre, Division of Mathematics, University of
Dundee, Dundee DD1 4HN, UK} \affiliation{Institute for Physics and
Technology, NASU, Donetsk 83114, Ukraine}
\author{M.~Chaplain}
\affiliation{The SIMBIOS Centre, Division of Mathematics, University of
Dundee, Dundee DD1 4HN, UK}
\author{V.~Levi}
\affiliation{Dallas Semiconductor, Dallas, Texas, US 75244 }

\begin{abstract}
We study the phenomenon of spatiotemporal stochastic resonance (STSR)
in a chain of diffusively coupled bistable oscillators. In particular,
we examine the situation in which the \textit{global} STSR response is
controlled by a \textit{locally applied signal} and reveal a wave front
propagation. In order to deepen the understanding of the system
dynamics, we introduce, on the time scale of STSR, the study of the
effective statistical renormalization of a generic lattice system.
Using this technique we provide a new criterion for STSR, and predict
and observe numerically a bifurcation-like behaviour that reflects the
difference between the most probable value of the local
quasi-equilibrium density and its mean value. Our results, tested with
a chain of nonlinear oscillators, appear to possess some universal
qualities and may stimulate a deeper search for more generic phenomena.
\end{abstract}

\date[Dated: ]{ \today}

\maketitle

Since the appearance of \cite{BSV81BSV83} the phenomenon of stochastic
resonance (SR) has become a popular field of research. A great deal of
experimental as well as theoretical and mathematical work has been
devoted to the study of the phenomenon in different systems (for
reviews, see \cite{RMP98,JSP93,chaos98,freid00}). There has been a
particular emphasis of its relevance and importance in biology and
medicine \cite{biomed} where noise in general, and SR in particular,
play a surprisingly constructive role \cite{noise}. The ability of SR
to generate order from disorder
\cite{RMP98,JSP93,chaos98,freid00,biomed,noise} is especially of
relevance in the context of pattern formation mechanisms that are
enhanced by noise \cite{pattern}. The discovery of an enhancement of
the effect by the coupling of nonlinear oscillators into an array
\cite{BSV85,Lind95Lind96} has brought new insight to studies of SR.
This effect in a wide sense is known as spatiotemporal SR (STSR)
\cite{chaos98}. An explanation of this effect has previously been
described as ``collective spatiotemporal motion'' and ``optimal
spatiotemporal synchronization''. In spite of much progress, the
precise kinetic details of such a synchronization remain without an
appropriate study.

We conjecture that among the kinetic details of STSR, wave front
propagation plays a prominent role (\cite{march96,march00} are good
background references in this regard). We investigate the related
problem of inducing and controlling global spatiotemporal order in a
chain of diffusively-coupled bistable oscillators by a \textit{locally
applied signal}. The ability to induce STSR throughout the chain by
applying a local signal to a small part of the chain is precisely
shown. This can be regarded as an element of a spatial signal
transmission, and possibly gives a new design freedom to modelling
biological and biomedical problems. We also investigate the effective
statistical renormalization of the steady states of a generic lattice
system \footnote{In contrast to \cite{pre93}, our approach is dynamic
by its nature and the spatially-homogeneous steady states appear as the
result of temporal evolution and provide more detailed results
\textit{cf.} \cite{pre93}.}. This renormalization reflects the
difference between the most probable and mean values of the local
quasi-equilibrium density which is a result of time averaging
\footnote{The point is likely applicable to allied effects
\cite{march01,noise}.}. This leads to a new observation that the
system, on the time scale of STSR, exhibits a bifurcation-like
behaviour. It also gives a criterion for the noise intensity depending
on the coupling in the chain. Both our results appear to have a certain
universal quality and may stimulate a deeper search for generic
phenomena, \textit{e.g.} in a chain of FitzHugh-Nagumo equations
\cite{kanamaru}.

Consider a chain of overdamped oscillators with diffusive coupling of
constant $K>0$ and a bistable on-site potential $V\left(
y\right)=-{my^2 }/2+{y^4}/4$. We assume that the system is influenced
both by external random noise of intensity $D$, which involves a set of
independent generalized Gaussian random processes $\left\{
{\xi_{n}(t)}\right\} $ with two characteristic cumulants, $
\left\langle {\xi_{n}(t)}\right\rangle =0 $ and $ \left\langle
{\xi_{m}(t)\xi_{n}(t^{\prime})}\right\rangle =\delta_{mn} \delta\left(
{t-t^{\prime}}\right) $, and a deterministic signal $S(t)$ applied
locally to a part of the chain, $ S_n^{\left( {M_k } \right)} \left( t
\right) = \left\{ {s(t)\quad {\text{if}}\quad n \in M_k ,\quad
{\text{and}}} \right.\quad 0\quad {\text{otherwise}}\} $. In what
follows we fix the particular and simplest form of the external signal,
$s(t)=A \cos(\omega t)$, together with dimensionless amplitude
$A=0.025$, and frequency $\omega=5\pi \cdot 10^{-5}$ that actually set
one of the timescales, $T_{\mathrm{s}} =2\pi\omega^{-1}=4\cdot10^4$.
The other characteristic timescales are: the relaxation time of the
chain to Gaussian fluctuations in the vicinity of one of its stable
steady states, $T_\mathrm{r}$ (in our case $T_\mathrm{r} \sim (2m)^{-1}
\sim 1$); the waiting time of the initial birth of an ``instanton'',
$T_\mathrm{K}$ (Kramers' time) \cite{BSV85} -- here we do not
explicitly consider this timescale but, fortunately, there are in-depth
studies of the problem \cite{march88,march98,march96} (\textit{a
posteriori} one can say that $T_\mathrm{K}$ is considerably shorter
than the principal timescale here) and also the timescale related to
any wave-front propagation in the chain, $T_\mathrm{w}$.

The corresponding chain stochastic differential equation (SDE) in
dimensionless variables has the form,
\begin{gather}\label{scde}
\dot{y}_{n}=K\Delta y_{n}-V^{\prime}\left(  {y_{n}}\right)
+\sqrt{2D}\xi _{n}(t)+S_{n}^{\left( {M_{k}}\right)  }\left( t\right)  ,
\end{gather}
with $n$ on the chain $\mathbb{N}$ in the integer lattice $\mathbb{Z}$,
$n \in \mathbb{N}=\{1,\ldots, k+1, \ldots, k+\mathrm{M}, \ldots,
\mathrm{N}\},\; M_{k} = \left\{{k+1, k+2, \ldots, k + \mathrm{M}}
\right\}, \;0 \leqslant \mathrm{M} \leqslant \mathrm{N}$; and $\Delta
y_{i} \equiv y_{i+1} - 2y_{i} + y_{i-1}$. Two different topologies are
possible for the chain: either the ends are connected or not connected.
We take the latter case with Neumann boundary condition.

The corresponding physical context of (\ref{scde}) is in the
Smoluchowski kinetics of a harmonically coupled chain of particles with
transverse displacements $\{y_n\}$. The underlying free energy
functional and the corresponding deterministic part of the dynamics,
without signal, are: ${
{\mathcal{F}}\left(y\right)=\sum_{\left(n\right)}\frac{1}{2}K{\left(
y_{n}-y_{n-1}\right)^2}
+V\left(y_{n}\right),\;\dot{y}_n=-\partial{\mathcal{F}}/\partial{y_n}}$.
This interpretation can considerably facilitate the understanding of
the dynamic behaviour of (\ref{scde}). Especially since even
diffusively-coupled, chain oscillators appear surprisingly difficult to
analyse consistently from a rigorous mathematical perspective
\cite{m-p98}.

We study a spatially discrete model because chains or array structures
are often of relevance in biology. Besides biology there are also crystal
lattices \cite{march98}. Moreover interesting dynamical effects exist in
discrete models that are not present in their continuous analogs -
\textit{e.g.} the propagation failure of travelling waves \cite{m-p98},
and breathers \cite{mp98}. Even in population dynamics it was recently
demonstrated that important effects due to the discrete nature of
organisms may be entirely missed by continuous models \cite{sci01}.

\begin{figure}[h]
\includegraphics[clip,angle=-90,width=8cm,totalheight=3cm]{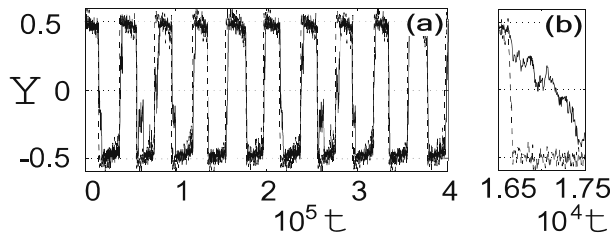}
\caption{(a) Dynamics of the collective variables $Y$ (solid) and $Y'$
(dash) with parameters $\textrm{N}=300$, $\mathrm{M}=50$, $k=125$,
$K=15$, $D=0.1$, and $m=0.25$; (b) Fine structure of the transition
region in (a) reveals a wave front propagation.} \label{f2}
\end{figure}

To introduce induced STSR phenomena, we first present a representative
numerical simulation of (\ref{scde}) that is performed longer than all
the characteristic timescales. We use time steps $0.01$ and $0.001$
that are considerably shorter then the principal timescale. Further
shortening of the step does not change the result.
\begin{figure}[h]
\includegraphics[clip,angle=-90,width=8cm,totalheight=4.5cm]{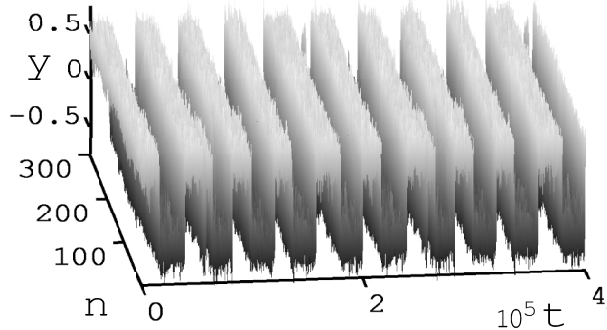}%
\caption{Locally induced global spatiotemporal pattern in the chain of
bistable oscillators; parameters are specified in Fig.~\ref{f2}. }
\label{f1}
\end{figure}
As shown in Fig.~\ref{f1} the system can indeed represent the
well-recognized phenomenon with a local signal applied to only $1/6$
part of the chain. The collective variables, $Y = \mathrm{N}^{-1}
\sum_{n=1}^\mathrm{N} y_n$ and $Y' = \mathrm{M}^{-1}
\sum_{n=k+1}^{k+\mathrm{M}} y_n$, also adequately and legibly reflect
the key features of the effect as shown in Fig.~\ref{f2}.

The generic features of STSR, \textit{i.e.} those related to a
diffusively coupled chain of bistable oscillators, have reasonable
prototypes going back as far as the pioneering paper \cite{BSV85}. The
main result from \cite{BSV85} is that, for transition times, only the
energy of the unstable instanton-like spatially inhomogeneous steady
state solutions, and not the total energy barrier for the chain, is of
importance \textit{cf.} \cite{fw84}. However, in fact, there has been
no discussion at all of the transition kinetics, which appear after the
birth of the ``instantons''.

The transition kinetics are most likely related to, but different from,
another interesting phenomenon - travelling waves \cite{m-p98}. It
is evident from the physical interpretation of (\ref{scde}), that a
travelling wave solution can appear only if the symmetry of the
bistable potential is broken and there is an energy gap between the two
stable states, and the energy flux is compensated for by dissipative
forces. This is not the case for the deterministic part of (\ref{scde})
with $A=0$. However, if $A\neq0$ then the external signal periodically
breaks the symmetry and creates an absolute minimum at one of the wells
of the underlying potential, and a travelling wave front can develop on
a sufficiently long timescale $T_\mathrm{s}$. The situation changes
favourably in the case of a chain because the translational symmetry is
broken. With the chain, starting from the physical interpretation
above, we can imagine an unstable instanton-like steady state solution
\cite{BSV85}. Its characteristic lifetime introduces a new time scale
that is required to be shorter than $T_\mathrm{s}$ for the continued
existence of the induced STSR effect. Our numerical simulations justify
\textit{a posteriori} that this condition is perfectly realizable.

Once the underlying kinetic mechanisms of STSR are understood, we can
explore other interesting features. Here we focus on one that has been
overlooked in previous studies, \textit{viz.} the shift of the stable
spatially homogeneous steady states.

Let us consider the specific lattice SDE
\begin{equation}
\label{slde}\dot{y}_{i}=K\Delta y_{i}-V^{\prime}\left( {y_{i}}\right)
+\sqrt{2D}\xi_{i}\left( t\right)  ,\quad i\in\mathbb{Z},
\end{equation}
and evolve the optimal Gaussian representation of this equation
(extending \cite{west83} to the case of a lattice system as a starting
point for further analysis; alternatively one can start with an
averaging principle \cite{fw84}),
\begin{equation}\label{gauss}
\dot{y}_{i} = K\Delta y_{i} - \left[ {a_{i} \left(  t \right) + b_{i}
\left( t \right) \delta y_{i} } \right] + \sqrt{2D} \xi_{i} \left( t
\right), \quad i \in \mathbb{Z},
\end{equation}
where $a_{i}$ and $b_{i}$ are obtained by a minimization procedure in
respect of the mean-square error functional $J =\left\langle {\left[
{V^{\prime} \left( {y_{i}} \right)  - \left( {a_{i} + b_{i} \delta
y_{i}} \right) }\right]^{2}} \right\rangle$ for all $i \in \mathbb{Z}$,
where $\delta y_{i} = y_{i} - \left\langle {y_{i}} \right\rangle$. The
representation (\ref{gauss}) of (\ref{slde}) is especially appropriate
since the \textit{ex ante} fluctuations are almost Gaussian near the
stable steady states and typical exit paths of (\ref{slde}). On the
timescale $T_\mathrm{s}$ the internal deterministic and random
oscillations are very fast and can be considered adiabatically
following with the external signal.

To actually obtain $a_{i}$ and $b_{i}$, which is an intractable problem
since it requires the exact solution of (\ref{slde}), we consider a
self-consistent approximation scheme combining the minimization of $J$
together with the solution of (\ref{gauss}). Thus we replace averaging
according to (\ref{slde}) by its Gaussian approximation according to
(\ref{gauss}), $ \bar y =\left\langle y \right\rangle
_{{\text{Gaussian}}} = \left\langle y \right\rangle $. As a result, we
obtain the self-consistent set
\begin{gather}
\dot{\bar y}_{i} = K \Delta \bar y_{i} - a_{i} \left(  t \right)  ,
\hfill\label{eff}\\
\delta \dot y_{i} = K \Delta \delta y_{i} - b_{i} \left(  t \right)
\delta y_{i} + \sqrt{2D} \xi_{i} \left(  t \right)  ,
\hfill\label{gfluct}\\
a_{i} \left(  t \right)  = \left\langle {V^{\prime}(y_{i} (t))}
\right\rangle, \quad b_{i} \left( t \right)  = \left\langle {\delta
y_{i} V^{\prime}(y_{i} (t))} \right\rangle \left\langle {\delta y_{i}
^{2} } \right\rangle^{-1}.\nonumber
\end{gather}
Further, using the Furutsu-Novikov formula \cite[and
references]{novikov64} we explicitly obtain $a_{i} = \left(
{3\mathfrak{K}_{2} (y_{i} ) - m} \right) \bar y_{i} + \bar y_{i} ^{3}
,\quad b_{i} = \left( {3\mathfrak{K}_{2} (y_{i} )} - m \right) + 3 \bar
y_{i} ^{2} $, where $\mathfrak{K}_{2} (y_{i} ) = \left\langle {y_{i}
^{2} } \right\rangle - {\bar y_{i}}^2 $ is the second cumulant. Thus
(\ref{eff}) takes the form $\dot{\bar y}_{i} = K \Delta \bar y_{i} +
\left(  {m - 3\mathfrak{K}_{2} (y_{i} )} \right) \bar y_{i} - \bar
y_{i} ^{3} , $ involving the effective potential function
$V_{{\text{eff}}}\;$, $V^{\prime}_{{\text{eff}}} \left( {\bar y_{i} }
\right)  = - \left( {m - 3\mathfrak{K}_{2} (y_{i} )} \right) \bar y_{i}
+ \bar y_{i}^{3}$. Since $\mathfrak{K}_{2} (y_{i} )\ge 0$,
$V_{\mathrm{eff}}\ge V $, always.

Lastly we consider the principal problem of an explicit calculation of
$\mathfrak{K}_{2} (y_{i} )$ and solve it under certain hypotheses.
Consider the spatial correlations, $\kappa_{mn} \left(  t \right) =
\left\langle {\delta y_{m} (t)\delta y_{n} (t)} \right\rangle $ and
suppose that the fluctuations tend to their steady state via a stage of
spatial homogenization with the ansatz $\kappa_{mn} \left( t \right)  =
\kappa_{m - n} \left(  t \right)  = \kappa_{r} \left(  t \right) $. The
known equilibrium solution of (\ref{slde}) with probability density
$\rho_\infty(y) \propto \exp[-\mathcal{F}(y)/D]$ has the property of
spatial homogeneity, but: Does the property still persists as time
evolves? To facilitate understanding, consider the process of the
formation of $\rho_\infty$ as a result of time averaging, and fix two
limiting cases related to low and high levels of the noise intensity.
As $D\downarrow 0$ the system spends most time in a potential trough,
occasionally passing from one to the other. Being in a trough it has
time to form a local quasi-equilibrium density that reflects the local
asymmetry of the underline potential. As $t\to +\infty$, a sum of the
local densities is formed to satisfy the global symmetry condition.
Under a high level of noise, the rate of passage from one trough to the
other is frequent enough in order to rapidly form the mean value $y=0$.
The qualitative picture described above is linked to the particular
time scale of the problem in question. Favourably for the ansatz, we
are interested in averaging over a time scale about $T_\mathrm{s}$ that
becomes apparent in the local quasi-equilibrium density. Test numerical
simulations are also essential in order to reinforce the ansatz.

\begin{figure}[t]
\includegraphics[clip,angle=-90,width=8cm,totalheight=4cm]{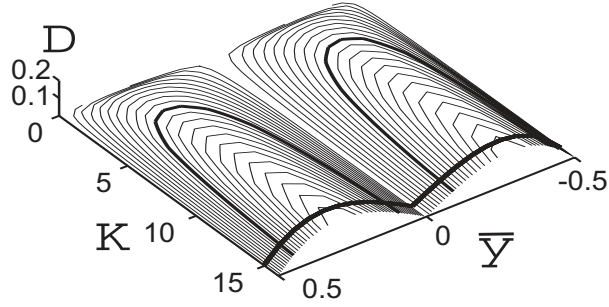}
\caption{Spatially-homogeneous steady states $\bar y \neq 0$ as points
of intersection of a characteristic surface $\bar y = \bar y (D,K)$ and
two coordinate planes of the control parameters ($D=0.1$ and $K=15$ are
marked). Instead of the two original steady states $\pm 0.5$ there are
now four for specific values of $D$ and $K$, and zero
otherwise.}\label{f3}
\end{figure}

\begin{figure}[tb]
\includegraphics[clip,angle=-90,width=8cm,totalheight=4cm]{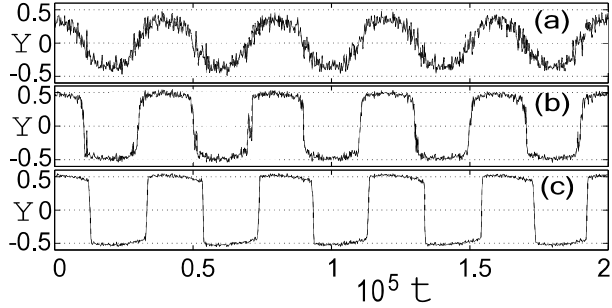}
\caption{Three output signals (collective variable $Y$,
$\mathrm{N}=\mathrm{M}=500$ ) with (a) $D=0.5$; (b) $D= 0.24$; and (c)
$D=0.1$; correspondingly, above, near, and below the characteristic
surface $\bar y=\bar y (D,K)$ at $K=15$ (see Fig.~\ref{f3}); effective
stable steady states: (a) $0$, and (c) $\pm0.43$. The difference
appears to be not only quantitative but also, more importantly,
qualitative.} \label{f4}
\end{figure}

Using (\ref{gfluct}) together with the Furutsu-Novikov formula, we
obtain a dynamical equation for $\kappa_{r} \left(  t \right) $: $\;
\dot\kappa_{r} = 2K\Delta\kappa_{r} - 2b(t)\kappa_{r} + 2D\delta_{r0}
,\quad r \in\mathbb{Z} $, -- and as a result, the equation for the
steady state correlation function, $ K\Delta\kappa_{r} - b\kappa_{r} +
D\delta_{r0}=0, \quad r \in\mathbb{Z}\;, $ with the natural asymptotic
conditions $ \mathop {\lim }\displaylimits_{r \to\pm\infty} \kappa_{r}
= 0 $; $\;b$ is still unknown. Substituting  $ \kappa_{r} = A \cdot t^{
- \left|  r \right|  } ,\quad t > 1, $ we obtain the set of algebraic
equations, corresponding to $r=0$ and $r\ge1\,$:  $\;\left[ {2K(1 - t^{
- 1} ) + b} \right]  A = D,\;\;\; t^{2} - 2(1 + {b}/{2K} )t + 1 = 0$.
The last equation has two different roots, $ t_{\pm}= (2K)^{-1}(2K + b
\pm\sqrt{b(4K + b)} ) $ , connected by the relation, $ t_{+} \cdot
t_{-} = 1 $. Since $ t \equiv t_{+} > 1 $, this means that $ t_{-} =
t^{ - 1} < 1 $. Therefore finally we obtain
\begin{gather*}
\kappa_{r} = \frac{D} {{\sqrt{b(4K + b)} }} \cdot\left[ {\frac{{2K + b
+ \sqrt{b(4K + b)} }} {{2K}}} \right]  ^{ - \left| r \right| } .
\end{gather*}
In particular, $\kappa_{0} = D/\sqrt{b(4K + b)}$. Further, we can use
$\kappa_{0}$ in combination with $V^{\prime}_{{\text{eff}}} \left(
{\bar y} \right) = 0$ to characterize the effective
spatially-homogeneous steady states. Excluding $b$ ($=3\kappa_0 - m
+3{\bar y}^2$) from this set, we arrive at two cases:
\begin{align*}
(a)& \quad \bar y = 0 ,&& (3\kappa_{0} - m)\left( {4K + 3\kappa_{0} -m}
\right)= \frac{{D^{2} }} {{\kappa_{0} ^{2} }},&&&\\
(b)& \quad  \bar y^2 = m - 3\kappa _0,&& (m - 3\kappa _0 )\left( {2K +
m - 3\kappa _0} \right) = \frac{{D^2 }} {{4\kappa _0 ^2 }}.&&&
\end{align*}
Observe first that while (a) always gives $\bar y =0$ and $\kappa_0\geq
m/3$, the set (b) has no real roots $\bar{y}$ ($\kappa_0>0$) for
a certain range of values of $D$ and $K$ (see Fig.~\ref{f3}).
In other words, the stable steady states disappear and only $\bar{y}=0$
remains in this range. We could claim that this occurs with a
bifurcation-like behaviour. To identify this observation, we carry out
numerical experiments on the system (\ref{scde}) varying the noise
intensity $D$ over a wide range. The response to the external signal
provides evidence of the effect as shown in Fig.~\ref{f4}, which is
enhanced further with larger $N$. In the case (c) the stochastic
resonance pattern, adjusted for the renormalization, is recognizable,
but in the case (a) only simple oscillations with the frequency of the
external signal are visible around $\bar y = 0$.

To further understanding of this observation, one can consider the
equilibrium probability density of (\ref{slde}). As $D\downarrow 0$,
$\rho_\infty(y)$ is concentrated at two spatially-homogeneous absolute
minima of the potential and forms the most improbable configurations
around unstable steady state $y=0$. The homogeneous equilibrium mean
value $\langle y\rangle_\infty = 0$ is fixed by the invariance of
$\rho_\infty( y)$ under the transformation $y\to -y$, and eventually
coincides with the unstable steady state. By increasing $D$ we can
change the sharpness of the equilibrium density profile, but not the
topology. It should be observed here that $\rho_\infty( y)$ represents
the average over all sample paths of (\ref{slde}). If we consider a
single sample path then the local density profile of (\ref{slde}) is
formed by time averaging, initially in the vicinity of an absolute
minimum of the potential reflecting the local potential asymmetry.
After a sufficiently long period the system with probability $1$ will
pass through the low-probability range to the vicinity of the other
absolute minimum of the potential, and so on. The rate of this process
rapidly increases with $D$ \cite{fw84}. The rate of development of a
difference between the most probable and mean values of the local
density is also increasing with $D$ since, within the same time
interval, more different states are realizable. It is remarkable that
the time interval related to the signal as well as the inherent nature
of SR is favourable to observe the mean values or, in other words, to
feel the effective potential. The collective variable $Y$, which is a
sort of site average, allows us to visualize this effect if the chain
is long enough.

Finally, if we set $D=0.1$ and $\sqrt{m}=0.5$, we obtain $\bar y
\approx\pm0.43$. This is close to the observed value obtained by direct
simulation of (\ref{scde}) (with $\mathrm{N}=\mathrm{M}=300$ and
$500$). The agreement improves with larger values of $\mathrm{N}$.
There also exists a pair of steady states close to $\bar y = 0$: $ \bar
y \approx \pm0.1$, -- but these are not clearly identified in the
simulation of STSR.

This work has been partially supported by research grants of the Royal
Society of London and the University of Dundee, Division of
Mathematics.


\begin{thebibliography}{99}
\bibitem{BSV81BSV83} R.~Benzi, A.~Sutera, and A.~Vulpiani, J. Phys. A
\textbf{14}, L453 (1981); R.~Benzi, G.~Parisi, A.~Sutera, and
A.~Vulpiani, SIAM J. Appl. Math. \textbf{43}, 565 (1983).
\bibitem{RMP98} L.~Gammaitoni, P.~H\"{a}nggi, P.~Jung, and
F.~Marchesoni, Rev. Mod. Phys. \textbf{70}, 223 (1998).
\bibitem{JSP93} F.~Moss, A.~Bulsara, and M.~Shlesinger (Eds.),
J. Stat. Phys. \textbf{70}, No. 1/2 (1993).
\bibitem{chaos98} M.~L\"{o}cher \textit{et al.},
Chaos \textbf{8}, 604 (1998).
\bibitem{freid00} M.I.~Freidlin, Physica D \textbf{137}, 333 (2000).
\bibitem{biomed} J.K.~Douglass \textit{et al.},
Nature (London) \textbf{365}, 337 (1993); K.~Wiesenfeld and F.~Moss,
\textit{ibid.} \textbf{373}, 33 (1995); J.E.~Levin and J.P.~Miller,
\textit{ibid.} \textbf{380}, 165 (1996); P.~Cordo \textit{et al.},
\textit{ibid.} \textbf{383}, 769 (1996); J.J.~Collins, \textit{ibid.}
\textbf{402}, 241 (1999); D.F.~Russell, L.A.~Wilkens, and F.~Moss,
\textit{ibid.} \textbf{402}, 291 (1999); F.~Jaramillo and
K.~Wiesenfeld, Nature Neurosci. \textbf{1}, 384 (1998); E.~Simonotto
\textit{et al.}, Phys. Rev. Lett. \textbf{78}, 1186 (1997); I.~Hidaka
\textit{et al.}, \textit{ibid.} \textbf{85}, 3740 (2000);
P.S.~Greenwood \textit{et al.}, Phys. Rev. Lett. \textbf{84}, 4773
(2000); T.~Mori and Sh.~Kai, \textit{ibid.} \textbf{88}, 218101 (2002);
B.~Spagnolo, \textit{et al.}, J. Phys.: Condens. Matter \textbf{14},
2247 (2002); Fluct. Noise Lett. \textbf{3}, L177 (2003); Physica A
\textbf{331}, 477 (2004).
\bibitem{noise} T.~Shinbrot and F.J.~Muzzio, Nature (London)
\textbf{410}, 251 (2001); L.~Glass, \textit{ibid.} \textbf{410}, 277
(2001); G.~Oster, \textit{ibid.} \textbf{417}, 25 (2002).
\bibitem{pattern} S.~K\'{a}d\'{a}r, J.~Wang, and K.~Showalter, Nature
(London) \textbf{391}, 770 (1998); J.M.G.~Vilar and J.M.~Rubi, Phys.
Rev. Lett. \textbf{78}, 2886 (1999).
\bibitem{BSV85} R.~Benzi, A.~Sutera, and A.~Vulpiani, J. Phys. A
\textbf{18}, 2239 (1985).
\bibitem{Lind95Lind96}J.F.~Lindner, \textit{et al.}, Phys. Rev. Lett.
\textbf{75}, 3 (1995); Phys. Rev. E \textbf{53}, 2081 (1996).
\bibitem{march96} F.~Marchesoni, L.~Gammaitoni, and A.R.~Bulsara,
Phys. Rev. Lett.  \textbf{76}, 2609 (1996).
\bibitem{march00} M.~L\"{o}cher, \textit{et al.},
Phys. Rev. E \textbf{61}, 4954 (2000).
\bibitem{pre93} F.J.~Alexander, S.~Habib, and A.~Kovner, Phys. Rev. E
\textbf{48}, 4284 (1993).
\bibitem{march01} G.~Costantini and F.~Marchesoni, Phys. Rev. Lett.
 \textbf{87}, 114102 (2001).
\bibitem{kanamaru} T.~Kanamaru, T.~Horita, and Y.~Okabe, Phys. Rev. E
\textbf{64}, 31908 (2001).
\bibitem{march88} P.~H\"{a}nggi, F.~Marchesoni, and P.~Sodano,
Phys. Rev. Lett. \textbf{60}, 2563 (1988).
\bibitem{march98} F.~Marchesoni, C.~Cattuto, and G.~Costantini,
Phys. Rev. B \textbf{57}, 7930 (1998).
\bibitem{m-p98} S.-N. Chow, J.~Mallet-Paret, and W.~Shen,
J. Differ. Equations \textbf{149}, 248 (1998).
\bibitem{mp98} M.~Peyrard, Physica D \textbf{119}, 184 (1998).
\bibitem{sci01} S.M.~Henson, \textit{et al.},
Science \textbf{294}, 602 (2001).
\bibitem{fw84} M.~Freidlin and A.~Wentzell , \textit{Random
Perturbations of Dynamical Systems} (Springer, Berlin, 1984).
\bibitem{west83} B.J.~West, G.~Rovner, and K.~Lindenberg,
J. Stat. Phys. \textbf{30}, 633 (1983).
\bibitem{novikov64} E.A.~Novikov, Sov. Phys. JETP \textbf{20}, 1290
(1965); A.~Samoletov, J. Stat. Phys. \textbf{96}, 1351 (1999).
\end{thebibliography}
\end{document}